\documentclass[]{aastex631}

\newcommand{\hetg}{{\small {\it Chandra}/HETG}}
\newcommand{\resolve}{{\small XRISM/{\it Resolve}}}
\newcommand{\kms}{km\,s$^{-1}$}
\newcommand{\NH}{$N_{\rm H}$}
\newcommand{\MBH}{$M_{\rm BH}$}
\newcommand{\NHtot}{$N_{\rm H}^{\rm tot}$}

\shorttitle{AASTeX v6.3.1 Sample article}
\shortauthors{Keshet \& Behar}
\graphicspath{{./}{figures/}}

\begin{document}

\title{ {\large Ionization Distributions in Outflows of Active Galaxies}\\ 
Universal Trends and Prospect of Future XRISM Observations}

\author{Noa Keshet}
\affiliation{Department of Physics, Technion, Haifa 32000, Israel}

\author{Ehud Behar}
\affiliation{Department of Physics, Technion, Haifa 32000, Israel}









\begin{abstract}

The physics behind the ionization structure of outflows from black holes is yet to be fully understood. Using archival observations with the \hetg\ gratings over the past two decades, we measured an absorption measure distribution for a sample of outflows in nine active galaxies (AGNs), namely the dependence of outflow column density, \NH, on the ionization parameter, $\xi$. The slope of $\log$\NH\ vs. $\log\xi$ is found to be between 0.00 and 0.72. We find an anti-correlation between the log of total column density of the outflow and the log of AGN luminosity, and none with the black hole mass and accretion efficiency. A major improvement in the diagnostics of AGN outflows will potentially occur with the launch of the \resolve\ spectrometer. We study the ability of \textit{Resolve} to reveal the outflow ionization structure by constructing the absorption measure distribution from simulated \textit{Resolve} spectra, utilizing its superior resolution and effective area. \textit{Resolve} constrains the column density as well as HETG, but with much shorter observations. 

\end{abstract}



\section{Introduction} \label{sec:intro}

Outflows are prevalent in Active Galactic Nuclei (AGNs). In the X-rays, one can observationally categorize three types of outflows. First are the relatively slow, $\sim 100$\,\kms , outflows detected through their numerous absorption lines in the soft X-ray spectra of Seyfert galaxies \citep[e.g.,][]{Kaastra2000}. Second, are the ultra-fast outflows ($\sim 0.1c$) detected mostly via highly blue-shifted Fe-K lines \citep{Tombesi2010, Nardini2015}. 
Third, are the high column-density outflows that almost completely obscure the soft X-ray AGN emission \citep{Kaastra2014}. The present work will focus on the Seyfert outflows and on their unique property of a broad distribution of ionization, which has not been identified in the other two types.

Seyfert outflows are known for their broad ionization distribution, from the most highly charge species to neutral and near neutral species \citep{Sako2001, Behar2001}. The ionization of a steady-state photo-ionized outflow can be described by its ionization parameter, which represents the balance between ionizing photons and recombining electrons. In the X-rays, one usually uses

\begin{equation}
\xi = \frac{L}{nr^2}
\end{equation}

\noindent 
where $L$ is the ionizing luminosity between 1 -- 1000\,Ryd, $n$ denotes the hydrogen number density, and $r$ is the distance of the absorber from the ionizing source at the AGN center.
$\xi$ therefore has dimensions, in cgs units that is erg\,s$^{-1}$cm. Seyfert outflows span a few orders of magnitude in $\xi$; henceforth we will refer to $-1 < \log \xi < 4$ in these units.

Previous works found several discrete ionization components in Seyfert outflows and discussed whether they represent a continuous distribution.
\citet{Blustin2005} published a survey of 23 AGNs to obtain insights into the absorbers and found that most outflows have multiple ionization phases, but there was no conclusion regarding whether the ionization structure is discrete or continuous. The average number of ionization components in their sample is two, but two AGNs with high quality spectra were modeled with three components. 
\citet{Detmers2011} studied the outflow of Mrk\,509 with a model that spans four orders of magnitude in $\xi$, but preferred discrete $\xi$ components over a continuous distribution. 
\citet{McKernan2007} analyzed X-ray spectra of 10 Seyfert outflows. 
Their best fits had two or more $\xi$-components, and five of the sources were modeled with three or more components. \citet{Laha2014} fitted \textit{XMM-Newton}/RGS spectra of 17 Seyfert outflows using 1 -- 3 $\xi$-components for each one, to obtain an ionization distribution common to all Seyfert outflows.

The column density \NH =$\int n dr$ of the outflow distributed over different $\xi$ values is termed the Absorption Measure Distribution \citep[AMD,][]{Holczer2007}:

\begin{equation}
    AMD = \frac{dN_{\rm H}}{d\log \xi}
\end{equation}

\noindent Since the AMD involves $n$ and $r$, its measurement can be invaluable for obtaining the density profile $n(r)$ of the outflow \citep{Behar2009}, which is not otherwise accessible via absorption measurements. This profile, in turn, may hold the crucial hint to the launching mechanism of Seyfert outflows, whether magnetic, radiative, or thermal (or a combination thereof). Different launching mechanisms could produce different AMDs, to be compared with the observed ones. For example by parameterizing the AMD as a power-law, with a slope $a$.
Magnetohydrodynamic (MHD) accretion disk winds predict an AMD slope that depends on the scaling of the magnetic field with radius (along the line of sight) $B\sim r^{q-2}$, leading to

\begin{equation}
AMD\sim \xi ^a = \xi^\frac{2q-2}{1-2q}
\end{equation}

\noindent or $q= (2+a)/(2+2a)$  \citep{Fukumura2010}.
Another physical effect is radiation compression of a hydro-statically held slab of photo-ionized gas \citep{Stern2014} which results in a broad and flat ($a\approx 0$) AMD. As the radiation is absorbed in the slab, its pressure drops, thus the gas pressure increases. Since the temperature also drops (less radiative heating) the density rises sharply, hence compression. Finally, numerical models of thermally driven winds have been able to produce AMDs that are flat, at least above $\log \xi =2$ \citep[][Fig.\,13 therein]{Waters2021}. 

In the present work, we model the spectra of a sample of nine Seyfert galaxies, using the \hetg\ spectrometer, seeking patterns in their outflow AMDs, and aiming to find a universal AMD profile. The specific goal is to study AMDs with commonly available spectral analysis tools, unlike the custom method of \citep[][]{Holczer2007}. Our secondary objective is to assess the abilities of the future calorimeter spectrometer \resolve\ \citep{Tashiro2020} for studying outflow AMDs. We examine advantages of the superior sensitivity of \resolve\ compared to \hetg. A comprehensive view of AGN outflows with the \textit{Hitomi} calorimeter, similar to that of \resolve\ was published by \citet{Kaastra2014white}. 

\section{Method} 
\label{sec:method}

\subsection{Sample and Data}
In order to study the AMD of AGN outflows, we selected all of the observations in the \hetg\ archive with visibly detectable absorption lines that can be ascribed to outflows. 
The targets and observations used in the present paper are detailed in Table \ref{tab:observations}. For each object, we used only observations from the same year in order to minimize variability of the absorber. We checked one such example, the NGC\,5548 HETG observations from 2000, 2002 and 2019, all with relatively high photon counts. All AGNs in the sample have a \hetg\ exposure of at least 190\,ks and at least 160,000 source counts. The rightmost column in Table \ref{tab:observations} lists previously published works  with multi-$\xi$ models for these observations, and for each target. Despite the variety of approaches, the models share some common attributes, as follows. They include 
$2 - 6\ \xi$-components \citep[e.g.,][]{Netzer2003}. Most of the models identify at least two kinematic components \citep[e.g.,][]{2005A&A...432..453S}; half of the objects have a slow outflow with velocity $100 - 500$ \kms\, and a fast one with velocity $>1000$ \kms\ \citep[e.g.,][]{Silva2016}. Out of these four objects with fast outflows, three are only apparent in the high-$\xi$ components. Two objects have no high-$\xi$ components nor fast outflows \citep{Gupta2013, Detmers2011}. In contrast with this variety of models, the present analysis uses a fixed set of ionization components, in order to obtain a uniform AMD structure.

\begin{deluxetable}{lccccccc}
\tabletypesize{\scriptsize}
\tablewidth{0pt}
\tablecaption{\hetg\ Observations used in The Present Work. }
\label{tab:observations}
\tablehead{
\colhead{AGN} & \colhead{Observations ID} & \colhead{Start Date}
& \colhead{Exposure} & \colhead{Total Exposure } & \colhead{ Total Counts}
& \colhead{Previous \small{HETG}}\\
\colhead{}   & \colhead{} & \colhead{}
& \colhead{[s]} & \colhead{[s]} & \colhead{1.5-20 \AA} & \colhead{Absorber Analysis}
}
\startdata
{        }     & 8452 & 2006 October 9   & 20755  & {}     & {}     & {}\\
{        }     & 7282 & 2006 October 10  & 41455  & {}     & {}     & {}\\
{NGC\,3516}    & 8451 & 2006 October 11  & 47567  & 190837 & 166320 & {\citet{Holczer2012}}\\
{        }     & 8450 & 2006 October 12  & 38550  & {}     & {}     & {}\\
{        }     & 7281 & 2006 October 14  & 42510  & {}     & {}     & {}\\
\hline
{        }     & 373  & 2000 January 20  & 56435  & {}     & {}     & {}\\
{        }     & 2090 & 2001 February 24 & 165659 & {}     & {}      & {\citet{Netzer2003}}\\
{NGC\,3783}    & 2091 & 2001 February 27 & 168961 & 889577 & 882032  & {\citet{Krongold2005}}\\
{        }     & 2092 & 2001 March 10    & 165454 & {}     & {}      & {\citet{Holczer2007}}\\
{        }     & 2093 & 2001 March 31    & 166885 & {}     & {}      & {\citet{Scott2014}}\\
{        }     & 2094 & 2001 June 26     & 166183 & {}     & {}      & {}\\
\hline
{        }     & 4760 & 2004 May 19      & 169898 & {}     & {}      & {}\\
{MCG -6-30-15} & 4761 & 2004 May 21      & 156641 & 524245 & 506528  & {\citet{Holczer2010}}\\
{        }     & 4759 & 2004 May 24      & 158539 & {}     & {}      & {}\\
{        }     & 4762 & 2004 May 27      & 39167  & {}     & {}     & {}\\
\hline
{        }     & 9151 & 2008 April 21    & 99980  & {}     & {}     & {}\\
{Ark 564 }     & 9899 & 2008 August 26   & 85000  & 347489 & 249462 & {\citet{Gupta2013}}\\
{        }     & 9898 & 2008 September 4 & 99772  & {}     & {}     & {}\\
{        }     & 10575& 2008 September 6 & 62737  & {}     & {}     & {}\\
\hline
{Mrk 509 }     & 13864& 2012 September 4 & 169939 & 268994 & 365281  & {\citet{Kaastra2014multi}}\\
{        }     & 13865& 2012 September 7 & 99055  & {}     & {}     & {}\\
\hline
{        }     & 2177 & 2001 August 26   & 59194  & {}     & {}     & {}\\
{        }     & 20070& 2017 June 6      & 91848  & {}     & {}     & {}\\
{IC 4329A}     & 19744& 2017 June 12     & 12380  & 233285 & 527009 & {\citet{Mehdipour2018}}\\
{        }     & 20095& 2017 June 13     & 33293  & {}     & {}     & {}\\
{        }     & 20096& 2017 June 14     & 19760  & {}     & {}     & {}\\
{        }     & 20097& 2017 June 17     & 16810  & {}     & {}     & {}\\
\hline
{        }     & 10777& 2008 November 6  & 27692  & {}     & {}     & {}\\
{        }     & 10775& 2008 November 8  & 30835  & {}     & {}     & {}\\
{        }     & 10403& 2008 November 9  & 37777  & {}     & {}     & {}\\
{        }     & 10776& 2008 November 11 & 24972  & {}     & {}     & {}\\
{        }     & 10778& 2008 November 11 & 33965  & {}     & {}     & {}\\
{NGC\,4051}    & 10404& 2008 November 12 & 19990  & 311092 & 245216 & {\citet{Lobban2011}}\\
{        }     & 10801& 2008 November 13 & 26077  & {}     & {}     & {\citet{King2012}}\\
{        }     & 10779& 2008 November 20 & 27312  & {}     & {}     & {}\\
{        }     & 10780& 2008 November 25 & 26055  & {}     & {}     & {}\\
{        }     & 10781& 2008 November 26 & 23940  & {}     & {}     & {}\\
{        }     & 10782& 2008 November 29 & 23315  & {}     & {}     & {}\\
{        }     & 10824& 2008 November 30 & 9162   & {}     & {}     & {}\\
\hline
{NGC\,4151}    & 3480 & 2002 May 7       & 90822  & 244024 & 240841 & {\citet{2005ApJ...633..693K}}\\
{        }     & 3052 & 2002 May 9       & 153202 & {}     & {}      & {}\\
\hline
{         }    & 837  & 2000 February 5  & 81957  & {}     & {}      & {}\\
{         }    & 3046 & 2002 January 16  & 152098 & {}     & {}      & {}\\
{NGC\,5548}    & 21846& 2019 May 5       & 29673  & 401808 & 225460  & {\citet{Steenbrugge2005B}}\\
{         }    & 22207& 2019 June 18     & 50307  & {}     & {}      & {}\\
{         }    & 21694& 2019 August 9    & 60528  & {}     & {}      & {}\\
{         }    & 22681& 2019 August 10   & 27245  & {}     & {}      & {}\\
\enddata
\end{deluxetable}

Table \ref{tab:parameters} lists the physical parameters of the AGNs. 
The unabsorbed X-ray luminosity $L_{\rm X}$ in the 0.5\,- 10\,keV band is measured from the present data, while the distance and black-hole mass \MBH\ are taken from the literature.
It can be seen that all AGNs are at low $z \le 0.034$, yet \MBH\ spans two orders of magnitude between $1.6\times10^6\,M_{\odot}- 1.4\times10^8\,M_{\odot}$, as does the accretion efficiency $0.003 \le L_{\rm X}/L_{\text{Edd}} \le 0.3$. Therefore, if the outflow AMD depends on any of these, we expect to be able to find hints to that dependence in our sample.

\begin{deluxetable}{lcccccccc}
\tabletypesize{\scriptsize}
\tablewidth{0pt}
\tablecaption{AGN Physical Parameters}
\label{tab:parameters}
\tablehead{
\colhead{AGN} & \colhead{z}
& \colhead{Distance\tablenotemark{\tiny{a}}}
& \colhead{Neutral \NH \tablenotemark{\tiny{b}}}
& \colhead{\MBH \tablenotemark{\tiny{c}}} 
& \colhead{$L_{\rm X}$\tablenotemark{\tiny{d}}} 
& \colhead{$ L_{\rm X}/L_{\text{Edd}}$ \tablenotemark{\tiny{e}}}\\
\colhead{} & \colhead{} & \colhead{[Mpc]} 
& \colhead{$10^{20}$[cm$^{-2}$]}
& \colhead{$[M_{\odot}]$} 
& \colhead{$10^{44}$[erg\,s$^{-1}]$} 
& \colhead{} 
}
\startdata
{NGC\,3516}    & 0.009 & 34.4  & 3.12 & $4.27 \pm 1.46\times10^7$
               & 0.132 & $0.0025 \pm 0.0008$ \\
{NGC\,3783}    & 0.010 & 42.0  & 10.1 & $2.98 \pm 0.54\times10^7$
               & 0.230 & $0.006 \pm 0.001$ \\
{MCG -6-30-15}& 0.008 & 33.3   & 4.1  & $1.6 \pm 0.4\times10^6$
               & 0.118 & $0.06 \pm 0.01$ \\
{Ark 564 }     & 0.025 & 105.8 & 6.4  & $2.61 \pm 0.26\times10^6$
               & 0.904 & $0.27 \pm 0.03$ \\
{Mrk 509 }     & 0.034 & 147.5 & 4.44 & $1.43 \pm 0.12\times10^8$
               & 2.409 & $0.013 \pm 0.001$ \\
{IC 4329A}     & 0.016 & 68.7  & 4.42 & $9.9 \pm 14.86 \times10^6$ \tablenotemark{\tiny{f}}
               & 1.350 & $0.1 \pm 0.2$ \\
{NGC\,4051}    & 0.002 & 10.1  & 1.35 & $1.91 \pm 0.78\times10^6$
               & 0.007 & $0.003 \pm 0.001$ \\
{NGC\,4151}    & 0.003 & 14.2  & 2.1  & $1.33 \pm 0.46\times10^7$
               & 0.105 & $0.006 \pm 0.002$ \\
{NGC\,5548}    & 0.017 & 73.1    & 1.55 & $6.71 \pm 0.26\times10^7$                                                  & 0.309 & $0.0037 \pm 0.0001$
\enddata
\tablenotetext{a}{Wright (2006) http://www.astro.ucla.edu/ Ewright/CosmoCalc.html}
\tablenotetext{b}{HEASARC $N_H$ TOOL \citep{NHtool}} 
\tablenotetext{c}{Masses for NGC\,3516, NGC\,3783, Mrk\,509, IC\,4329A, NGC\,4051, NGC\,4151 and NGC\,5548 were taken from \citet{Peterson2004}.
Mass of MCG\,-6-30-15 was taken from \citet{Bentz2016}. }
\tablenotetext{d}{Mean unabsorbed observed X-ray luminosity in the 0.5-10 keV band}
\tablenotetext{e}{Errors for the luminosity ratio were calculated based on mass errors}
\tablenotetext{f}{\citet{Calle2010} report $1.2\times10^8[M_{\odot}]$.}
\end{deluxetable}

 The simultaneous fitting of all spectra of each target assumes the absorber did not vary (or it yields results for a mean outflow of that target) during the temporal period of observations. Since the overwhelming majority of observations for a given target were obtained within a year, or even much less, this seems to be a reasonable assumption for Seyfert outflows, also supported by the previous analyses. The unresolved (narrow) absorption lines  suggest the absorber is relatively far from the nucleus, and would not vary over such short time scales. 
However, there are reports of absorber variability \citep{Krongold2005, Krongold2007}.

\subsection{Spectral Model}
    In each observation, first diffraction order ($\pm1$) of both the HEG and the MEG gratings were used. The total exposure time for all targets is at least 190\,ks, and the total photon counts between 1.5 and 20 \AA\ are greater than 166320 for all objects. We prepared different XSTAR \citep{Bautista2001} tables for each target, with the appropriate power-law spectral index $\Gamma$, a soft excess component when needed, and using solar abundances. We use a turbulent velocity of 100\,\kms\ , 
    except for the broad lines of NGC\,3516 \citep{Holczer2012} and the $\log \xi = 4$ component of MCG\,-6-30-15 \citep{Holczer2010}. The grid includes 7 logarithmic steps between $\log$\NH = $10^{18} - 10^{24}$, and 9 logarithmic steps in $\log\xi$ from -1 to 4.       

We fitted the spectra using the the Xspec implementation of C-statistic \citep{Cash1976}, since some of the bins in the data contain only few photons. The above tables were sufficient to well-fit (C-stat/dof $\sim1$) all spectra with components describing the continuum, and several absorption ionization components \NH$(\xi)$.

 We fitted the spectra of each target between 1.5-20 \AA\ using Xspec \citep{Arnaud1996}. 
 The continua are modeled as a power-law with a soft excess (when needed) that rises above the power-law around 15\,\AA . 
 This soft-excess is commonly modeled as a black-body component, which is satisfactory for our purpose of characterizing the absorber. 
The neutral galactic absorption was also taken into account with a fixed absorption component (Table\,\ref{tab:parameters}).
We modeled the absorber with six components at fixed $\log\xi_i$ values of -1, 0, 1, 2, 3, \text{ and} 4. The exception is NGC\,3783, where we added two more components at 0.5 and 2.5, required by its superb spectrum. 

The velocity of each $\xi_i$-component was fitted individually.
The outflows of NGC\,3516 and MCG\,-6-30-15 have more complex kinematic structure.
NGC\,3516 requires at least two kinematic components. One between -$650  -$ -$50$\,\kms\, and the second at $-1650\pm50$. Both kinematic components have all six of the $\xi_i$-components. The turbulent velocity in each outflow was kept fixed for all $\xi_i$-components.
The outflow of MCG\,-6-30-15 also has two kinematic components; the $\log\xi=4$ component is at $-1800\pm80$\,\kms, while the lower-$\xi$-components are between $-350- -130$\,\kms.


Finally we allowed the fit to include the following narrow emission lines - 1.78\AA\, (Fe Ly$\alpha$), 1.87\AA\, (Fe$^{+24}\,f$) , 1.94\AA\, (Fe K$\alpha$), 13.45\AA\, (Ne$^{+8}\,r$), 13.70\AA\, (Ne$^{+8}\,f$).
NGC 4151 included the following emission features as well:  7.13\,\AA\ (Si K$\alpha$),
9.17\,\AA\ (Mg$^{+10}\,r$),
9.31\,\AA\ (Mg$^{+10}\,f$),
12.13\,\AA\ (Ne$^{+9}$ Ly$\alpha$), 
16.01\,\AA\ (O$^{+7}$ Ly$\beta$), 
16.8\AA\ (O$^{+6}$ RRC).



\subsection{AMD Reconstruction}
The best-fit column densities $N_{\rm H}(\xi _i)$ of each AGN are subsequently used to build the AMD, by plotting $\log{N_H}$ as a function of $\log\xi_i$. For each target, we obtain the AMD slope by fitting a linear regression in log space to  $\log N_{\rm H}$ vs. $\log\xi_i$, taking into account $N_{\rm H}(\xi _i)$ uncertainties, which are calculated by Xspec with the standard 90\% confidence. For NGC\,3516 we built the AMD based on the values of the slower outflow velocity. 
The uncertainties on each $N_{\rm H}(\xi _i)$ for this purpose are taken as a mean of the lower and upper statistical uncertainties extracted from the spectral fit. The total column density, $N_{\rm H}^{\rm tot} = \Sigma_i N_{\rm H}(\xi _i)$, is obtained for each AGN. The reported uncertainty of \NHtot reflects the maximal lower and upper limits.

We subsequently take the best-fit \hetg\ model to simulate \resolve\ spectra using its anticipated response matrix \citep{Ishisaki2018}. We simulate a standard exposure time of 100 ks for all targets. For Ark\,564, Mrk\,509, IC\,4329A, NGC\,4051, NGC\,4151 and NGC\,5548, the highest $\xi$ component, $\log\xi=4$ is not well constrained with \hetg . 
Since the simulation randomly draws photons, it is meaningless to simulate such low column densities that the spectrometer cannot detect. Thus, in the \resolve\ simulations of these targets we take the HETG upper limit. In order to get an idea of the ability of \resolve\ to constrain the AMD, we then fitted the same model to the simulated spectra, followed by constructing an AMD for each object. Next, we carried out the linear regression and calculated \NHtot\, to be compared with the \hetg\ results.

\section{Results} \label{sec:results}

\begin{figure}
    \centering
    \includegraphics[width=\textwidth]{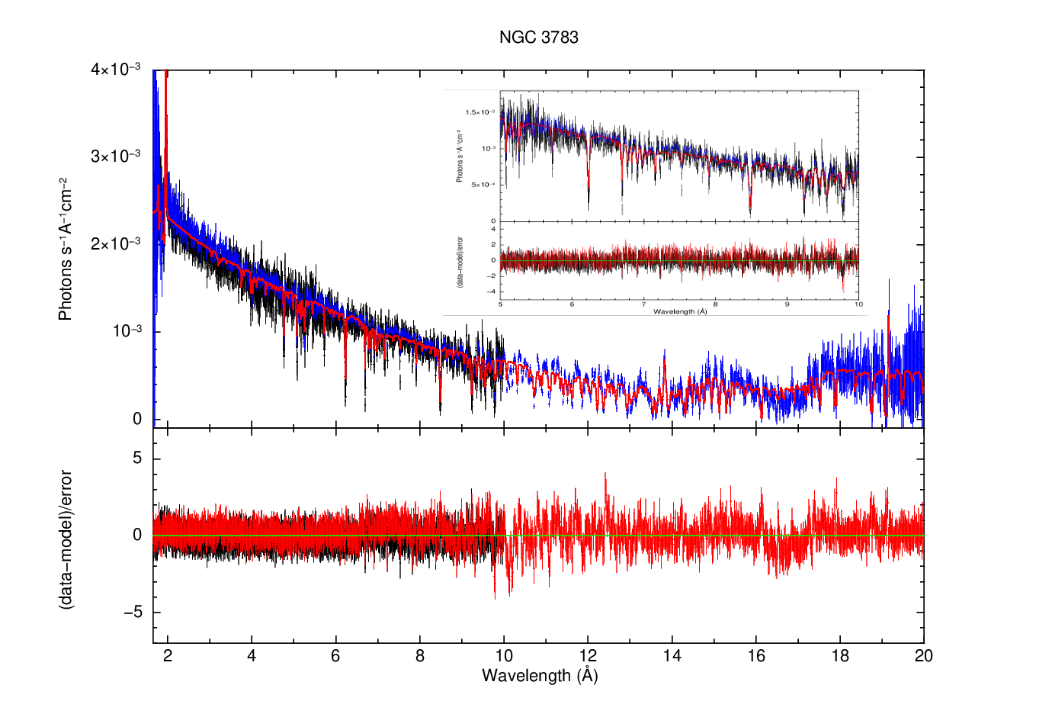}
    \caption{HETG, HEG (black) and MEG (blue) data and fitted model (red) of NGC\,3783, with model residuals (bottom panel). Insert shows zoom-in to the $5-10$ \AA\, region.}
    \label{fig:3783spectra}
\end{figure}

\begin{figure}
    \centering
    \includegraphics[width=\textwidth]{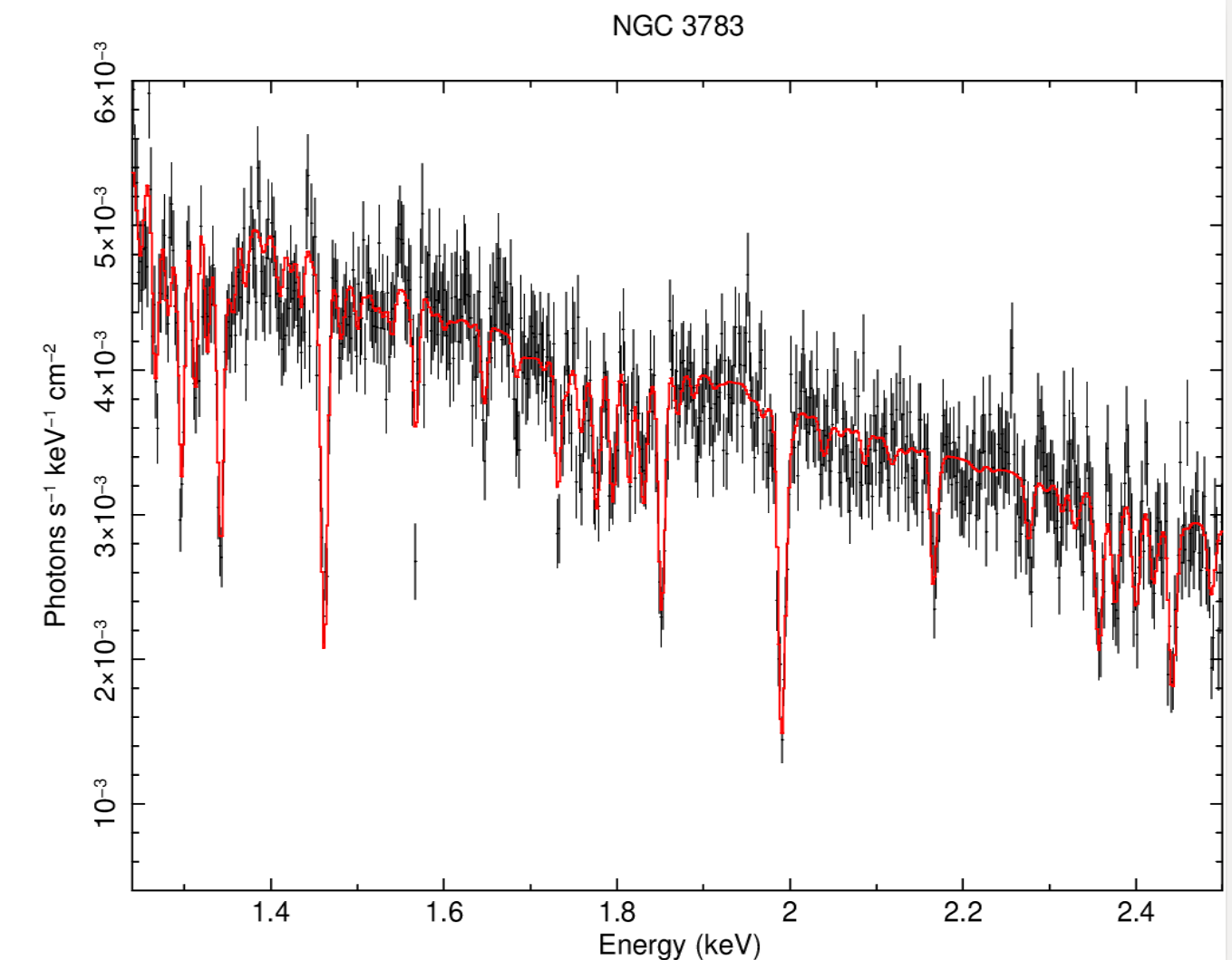}
    \caption{Simulated \textit{Resolve} spectrum (black) and fitted model (red) of NGC\,3783.}
    \label{fig:3783Xrism}
\end{figure}

An example of the HETG spectra and fitted model are shown for NGC\,3783, in Figure\,\ref{fig:3783spectra}. The residuals of the fit, 
visually demonstrate the adequacy of the model. The Figure includes a zoom-in that allows a closer view of some of the absorption lines. Figure\,\ref{fig:3783Xrism} shows the same part of the spectrum in the \resolve\ simulation, plotted in energy instead of wavelength to best demonstrate the performance of the future spectrometer.

\subsection{Column Densities}
The best-fit column densities for each $\xi_i$-component of each target, as well as \NHtot, are listed in Table\,\ref{tab:columns}. The best-fit continuum parameters and cstat/dof values are listed in Table\,\ref{tab:continuum}.
As expected, the best-fit column densities of \hetg\ observations and \resolve\ simulations agree to within 60\%, and are consistent within the uncertainties. 
However, the fractional uncertainties of the \resolve\ fits are generally smaller by up to a factor 4, and especially in the highest $\xi$-components. 
The improvement of \resolve\ is most noticeable in the $\log\xi = 4$ component of NGC\,4151 and NGC\,5548.
We also examined the difference between \textit{Resolve} simulations when taking the best-fit HETG value, versus using the upper limit for $\log\xi=4$. The results are very similar for Ark\,564 and NGC\,4051, where the column density of the $\log\xi = 4$ component is not constrained neither in HETG nor in \textit{Resolve}. For Mrk\,509 and IC\,4329A the column of the $\log\xi = 4$ component is again not constrained when the original column density value is used, as opposed to the upper limit value. 

\begin{deluxetable}{lccccccccccc}
\tabletypesize{\scriptsize}
\tablewidth{0pt}
\tablecaption{Best-fit Column Densities $N_{\rm H}(\xi _i)$ \tablenotemark{\footnotesize{a}} of the various $\xi_i$-Components, for HETG and \textit{Resolve}
\label{tab:columns}}
\tablehead{
\colhead{AGN} & \colhead{$\log\xi _i$ \tablenotemark{\footnotesize{b}}}
& \colhead{-1.0}
& \colhead{0.0} & \colhead{0.5} 
& \colhead{1.0} & \colhead{2.0} & \colhead{2.5}
& \colhead{3.0} & \colhead{4.0}
& \colhead{$\Sigma_i N_{\rm H}(\xi _i)$}
& \colhead{Outflow Velocity Range\tablenotemark{\footnotesize{c}}}
}
\startdata
{NGC\,3516} &    HETG         & $0.4^{+0.2}_{-0.1}$     
                              & $\leq0.02$
                              & --   
                              & $0.28^{+0.08}_{-0.07}$        
                              & $\leq0.02$ 
                              & --   
                              & $1.7^{+0.3}_{-0.2}$  
                              & $5.5^{+1.7}_{-1.3}$
                              & $7.9^{+2.0}_{-1.7}$
                              & $350-780$\\
{         } &    {  }         & $0.2^{+0.1}_{-0.1}$     
                              & $\leq0.01$
                              & --   
                              & $1.7^{+0.2}_{-0.2}$        
                              & $\leq0.02$ 
                              & --   
                              & $2.9^{+0.4}_{-0.3}$  
                              & $30^{+1.0}_{-1.0}$
                              & $34\pm1.7$
                              & $1670^{+60}_{-20}$\\        
{        } & \textit{Resolve} & $0.3^{+0.09}_{-0.07}$ 
                              & $\leq0.1$  
                              & --    
                              & $0.28^{+0.09}_{-0.07}$
                              & $\leq0.1$  
                              & --   
                              & $1.7^{+0.2}_{-0.2}$  
                              & $6.1^{+1.7}_{-1.3}$
                              & $8.5^{+2.3}_{-1.7}$
                              & $60-700$\\
{         } &    {  }         & $0.1^{+0.06}_{-0.07}$     
                              & $\leq0.2$
                              & --   
                              & $1.7^{+0.2}_{-0.2}$        
                              & $\leq0.1$ 
                              & --   
                              & $2.8^{+0.3}_{-0.3}$  
                              & $31.6^{+1.7}_{-1.6}$
                              & $36.3^{+2.6}_{-2.3}$
                              & $1680^{+30}_{-40}$\\                               
\hline
{NGC\,3783} &    HETG          & $1.4^{+0.2}_{-0.2}$
                              & $\leq0.02$ 
                              & $1.9^{+0.5}_{-0.5}$ 
                              & $1.7^{+0.2}_{-0.2}$
                              & $5.8^{+0.4}_{-0.4}$  
                              & $1.6^{+0.2}_{-0.2}$ 
                              & $29^{+2}_{-2}$  
                              & $21^{+9}_{-6}$
                              & $62^{+12}_{-9}$
                              & $230-300, 830-1440$\\
{        } & \textit{Resolve} & $1.6^{+0.2}_{-0.2}$  
                              & $\leq1.9$ 
                              & $0.7^{+1.8}_{-0.5}$ 
                              & $2.4^{+0.7}_{-0.7}$
                              & $5.9^{+1.1}_{-1.0}$  
                              & $1.8^{+0.6}_{-0.4}$ 
                              & $28^{+3}_{-3}$  
                              & $23^{+7}_{-6}$
                              & $63^{+17}_{-12}$
                              & $180-320, 540-1230$\\
\hline
{MCG -6-30-15}&    HETG       & $\leq0.3$ 
                              & $1.7^{+0.3}_{-0.3}$  
                              & --    
                              & $0.46^{+0.1}_{-0.09}$
                              & $0.6^{+0.1}_{-0.1}$  
                              & --   
                              & $2.5^{+0.2}_{-0.2}$   
                              & $19^{+2}_{-2}$
                              & $25^{+3}_{-3}$
                              & $130-350, 1820^{+70}_{-80}$\\
{        } & \textit{Resolve} & $\leq0.3$
                              & $1.8^{+0.4}_{-0.3}$  
                              & --    
                              & $0.4^{+0.1}_{-0.1}$
                              & $0.6^{+0.3}_{-0.2}$  
                              & --   
                              & $2.0^{+0.3}_{-0.3}$  
                              & $19^{+2}_{-2}$
                              & $24^{+4}_{-3}$
                              & $20-760, 1830^{+80}_{-80}$\\
\hline
{Ark 564}  &    HETG          & $\leq0.05$ 
                              & $0.3^{+0.1}_{-0.1}$ 
                              & --    
                              & $0.07^{+0.07}_{-0.05}$
                              & $0.8^{+0.2}_{-0.1}$   
                              & --   
                              & $0.5^{+0.3}_{-0.2}$ 
                              & $\leq1.2$
                              & $1.6^{+1.9}_{-0.4}$
                              & $60-280$\\
{        } & \textit{Resolve} & $\leq0.1$ 
                              & $0.6^{+0.2}_{-0.2}$ 
                              & --   
                              & $\leq0.02$
                              & $0.9^{+0.2}_{-0.2}$  
                              & --   
                              & $1.0^{+0.1}_{-0.3}$ 
                              & $\leq1.4$
                              & $2.7^{+2.1}_{-0.9}$
                              & $80-350$\\
\hline
{Mrk 509}  &    HETG          & $\leq0.02$ 
                              & $\leq0.1$ 
                              & --    
                              & $0.1^{+0.05}_{-0.07}$
                              & $1.2^{+0.1}_{-0.1}$ 
                              & --   
                              & $0.9^{+0.3}_{-0.7}$  
                              & $\leq17$
                              & $3^{+18}_{-2}$
                              & $10-600$\\
{        } & \textit{Resolve} & $\leq0.06$ 
                              & $\leq0.1$ 
                              & --    
                              & $0.16^{0.04}_{0.04}$
                              & $1.2^{+0.1}_{-0.1}$  
                              & --   
                              & $0.7^{+0.5}_{-0.4}$  
                              & $\leq13$
                              & $8^{+7}_{-7}$
                              & $60-770$\\
\hline
{IC 4329A} &    HETG          & $1.7^{+0.2}_{-0.1}$  
                              & $3.2^{+0.2}_{-0.2}$  
                              & --    
                              & $0.7^{+0.3}_{-0.2}$
                              & $1.2^{+0.1}_{-0.1}$ 
                              & --   
                              & $2.8^{+0.6}_{-0.6}$  
                              & $\leq6$
                              & $10^{+8}_{-1}$
                              & $50-420$\\
{        } & \textit{Resolve} & $1.8^{+0.1}_{-0.1}$   
                              & $3.2^{+0.1}_{-0.1}$
                              & --   
                              & $0.7^{+0.2}_{-0.2}$
                              & $1.3^{+0.1}_{-0.1}$ 
                              & --   
                              & $2.4^{+0.5}_{-0.4}$  
                              & $\leq11$
                              & $12^{+9}_{-4}$
                              & $60-360$\\
\hline
{NGC\,4051} &    HETG         & $\leq0.02$
                              & $\leq0.2$ 
                              & --    
                              & $0.11^{+0.04}_{-0.06}$
                              & $1.3^{+0.2}_{-0.2}$ 
                              & --   
                              & $1.9^{+0.4}_{-0.4}$  
                              & $\leq4.2$
                              & $3.4^{+5.0}_{-0.6}$
                              & $100-500$\\
{        } & \textit{Resolve} & $\leq0.1$ 
                              & $\leq0.2$
                              & --    
                              & $0.09^{+0.03}_{-0.08}$
                              & $1.3^{+0.2}_{-0.1}$ 
                              & --   
                              & $2.0^{+0.4}_{-0.4}$  
                              & $\leq10$
                              & $4^{+11}_{-1}$
                              & $190-600$\\
\hline
{NGC\,4151}  &    HETG        & $5.3^{+0.2}_{-0.2}$  
                              & $\leq0.01$ 
                              & --    
                              & $73^{+1}_{-1}$
                              & $\leq0.03$  
                              & --   
                              & $7.3^{+3.3}_{-2.6}$  
                              & $\leq25$
                              & $86^{+30}_{-4}$
                              & $40-200$\\
{        } & \textit{Resolve} & $5.1^{+0.2}_{-0.3}$  
                              & $\leq0.3$ 
                              & --    
                              & $73^{+6}_{-6}$
                              & $\leq0.2$  
                              & --   
                              & $8.5^{+0.2}_{-0.2}$  
                              & $23^{+10}_{-7}$
                              & $110^{+19}_{-11}$
                              & $10-200$\\
\hline
{NGC\,5548}  &    HETG        & $\leq0.01$  
                              & $\leq0.01$ 
                              & --    
                              & $0.2^{+0.09}_{-0.06}$
                              & $3.0^{+0.3}_{-0.3}$  
                              & --   
                              & $12^{+2}_{-2}$  
                              & $\leq105$
                              & $15^{+107}_{-2}$
                              & $40-950$\\
{        } & \textit{Resolve} & $\leq0.05$  
                              & $\leq0.7$ 
                              & --    
                              & $0.2^{+0.1}_{-0.1}$
                              & $2.8^{+0.4}_{-0.3}$  
                              & --   
                              & $12^{+2}_{-2}$  
                              & $110^{+38}_{-63}$
                              & $126^{+42}_{-65}$
                              & $30-1330$\\
\enddata
\tablenotetext{a}{in units of $10^{21}$\,cm$^{-2}$}
\tablenotetext{b}{$\xi _i$ in units of erg\,s$^{-1}$cm}
\tablenotetext{c}{$v$ in units of km/s, the range for all components with errors. The velocities are written without the minus sign for convenience.}
\end{deluxetable}

\begin{deluxetable}{lccccccc}
\tabletypesize{\scriptsize}
\tablewidth{0pt}
\tablecaption{Best-fit Continuum Parameters, for HETG and \textit{Resolve}}
\label{tab:continuum}
\tablehead{
\colhead{AGN} & \colhead{} 
& \colhead{$\Gamma$}
& \colhead{norm [photons/keV/cm$^2$/s]} & \colhead{kT [keV]} 
& \colhead{norm [$10^{-6}$ erg/s/cm$^2$]} & \colhead{cstat/dof} 
}
\startdata
{NGC\,3516}  &      HETG       & $1.55\pm0.01$                                                                                       & $0.0120\pm0.0001$ 
                               & $0.081\pm0.004$                                                      & $0.0022\pm0.0005$                                                    & 1.2675 \\
{         }  & \textit{Resolve}& $1.48\pm0.01$                                                                                       & $0.0104\pm0.0002$                                                                                   & $0.072\pm0.002$                                                                                     & $0.0037\pm0.0005$                                                                                   & 0.9978\\
\hline
{NGC\,3783}  &      HETG       & $1.75\pm0.01$                                                                                       & $0.0189\pm0.0004$                                                                                   & $0.097\pm0.002$                                                                                     & $0.0015\pm0.0001$                                                                                   & 1.6318 \\
{         }  & \textit{Resolve}& $1.75\pm0.02$                                                                                       & $0.0190\pm0.0005$                                                                                   & $0.098\pm0.004$                                                                                     & $0.0015\pm0.0004$                                                                                   & 1.0069 \\
\hline
{MCG\,-6-30-15}&    HETG       & $1.80\pm0.02$                                                                                       & $0.0132\pm0.0003$                                                                                   & $0.066\pm0.003$                                                                                     & $0.005\pm0.001$                                                                                     & 1.1097 \\
{            }&\textit{Resolve}& $1.79\pm0.02$                                                                                       & $0.0132\pm0.0003$                                                                                   & $0.067\pm0.002$                                                                                     & $0.005\pm0.001$                                                                                     & 1.0165 \\
\hline
{Ark\,564}   &      HETG       & $2.48\pm0.02$                                                                                       & $0.0164\pm0.0003$                                                                                   & $0.129\pm0.002$                                                                                     & $0.00073\pm0.00006$                                                                                 & 1.0785 \\
{        }   & \textit{Resolve}& $2.48\pm0.01$                                                                                       & $0.0165\pm0.0002$                                                                                   & $0.129\pm0.002$                                                                                     & $0.00079\pm0.00008$                                                                                 & 1.0524 \\
\hline
{Mrk\,509}   &      HETG       & $1.86\pm0.01$                                                                                       & $0.0178\pm0.0002$                                                                                   & $0.117\pm0.007$                                                                                     & $0.00029\pm0.00005$                                                                                 & 1.0710 \\
{        }   & \textit{Resolve}& $1.865\pm0.008$                                                                                     & $0.0179\pm0.0002$                                                                                   & $0.117\pm0.004$                                                                                     & $0.00033\pm0.00006$                                                                                 & 1.0101 \\
\hline
{IC\,4329A}  &      HETG       & $1.90\pm0.02$                                                                                       & $0.049\pm0.001$                                                                                     & --                                                                                                  & --                                                                                                  & 1.1547 \\
{         }  & \textit{Resolve}& $1.908\pm0.009$                                                                                     & $0.0495\pm0.0007$                                                                                   & --                                                                                                  & --                                                                                                  & 1.0152 \\
\hline
{NGC\,4051}  &      HETG       & $2.02\pm0.01$                                                                                       & $0.0103\pm0.0002$                                                                                   & $0.111\pm0.0.004$                                                                                   & $0.00041\pm0.00005$                                                                                 & 1.1662 \\
{         }  & \textit{Resolve}& $2.01\pm0.01$                                                                                       & $0.0102\pm0.0002$                                                                                   & $0.111\pm0.002$                                                                                     & $0.00041\pm0.00005$                                                                                 & 1.0372 \\
\hline
{NGC\,4151}  &      HETG       & $2.01\pm0.02$                                                                                       & $0.094\pm0.004$                                                                                     & --                                                                                                  & --                                                                                                  & 1.2215 \\
{         }  & \textit{Resolve}& $2.00\pm0.01$                                                                                       & $0.092\pm0.002$                                                                                     & --                                                                                                  & --                                                                                                  & 1.0130 \\
\hline
{NGC\,5548}  &      HETG       & $1.22\pm0.01$                                                                                       & $0.00512\pm0.00008$                                                                                 & --                                                                                                  & --                                                                                                  & 1.0633 \\
{         }  &      {   }      & $1.68\pm0.01$                                                                                       & $0.0070\pm0.0001$                                                                                   & --                                                                                                  & --                                                                                                  & 1.1464 \\
{         }  & \textit{Resolve}& $1.22\pm0.01$                                                                                       & $0.00520\pm0.00005$                                                                                 & --                                                                                                  & --                                                                                                  & 1.0100
\enddata
\end{deluxetable}

For the most part, our results of \NHtot and outflow velocities approximately agree with previous works, apart from the two outstanding following exceptions. 
\citet{Gupta2013} fitted the outflow of Ark\,564 with two $\xi$-components, both at a velocity of $\sim -100$\,\kms , at $\log \xi = 0.39\pm0.03$ and at $-0.99\pm0.13$. Their total columns at $\log N_H (\text{cm}^{-2}) = 20.94 \text{ and } 20.11$, respectively, are much lower than what we find, likely because \citet{Gupta2013} analyzed only spectra above $\sim9$\,\AA , therefore finding no high-$\xi$ components. 
The \textit{XMM-Newton}/RGS spectrum of the outflow of NGC\,4051 was fitted by \citet{Silva2016} with four ionization components: $\log \xi = 0.37\pm0.03, 2.60\pm 0.10, 2.99\pm0.03, \text{ and } 3.70\pm0.04$ corresponding to 
outflow velocities of, respectively, $-340\pm10, -530\pm10, -4260\pm60, \text{ and} -5770\pm30$~\kms . We do not identify the three high-velocity components in the HETG spectra. 
For NGC\,3783 we found two velocity ranges between the $\xi_i$-components, within the range of ionic velocities reported in  \citet{Kaspi2002}.

\subsection{AMDs} \label{subsec:AMD}
\begin{figure}
    \centering
    \includegraphics[width=\textwidth]{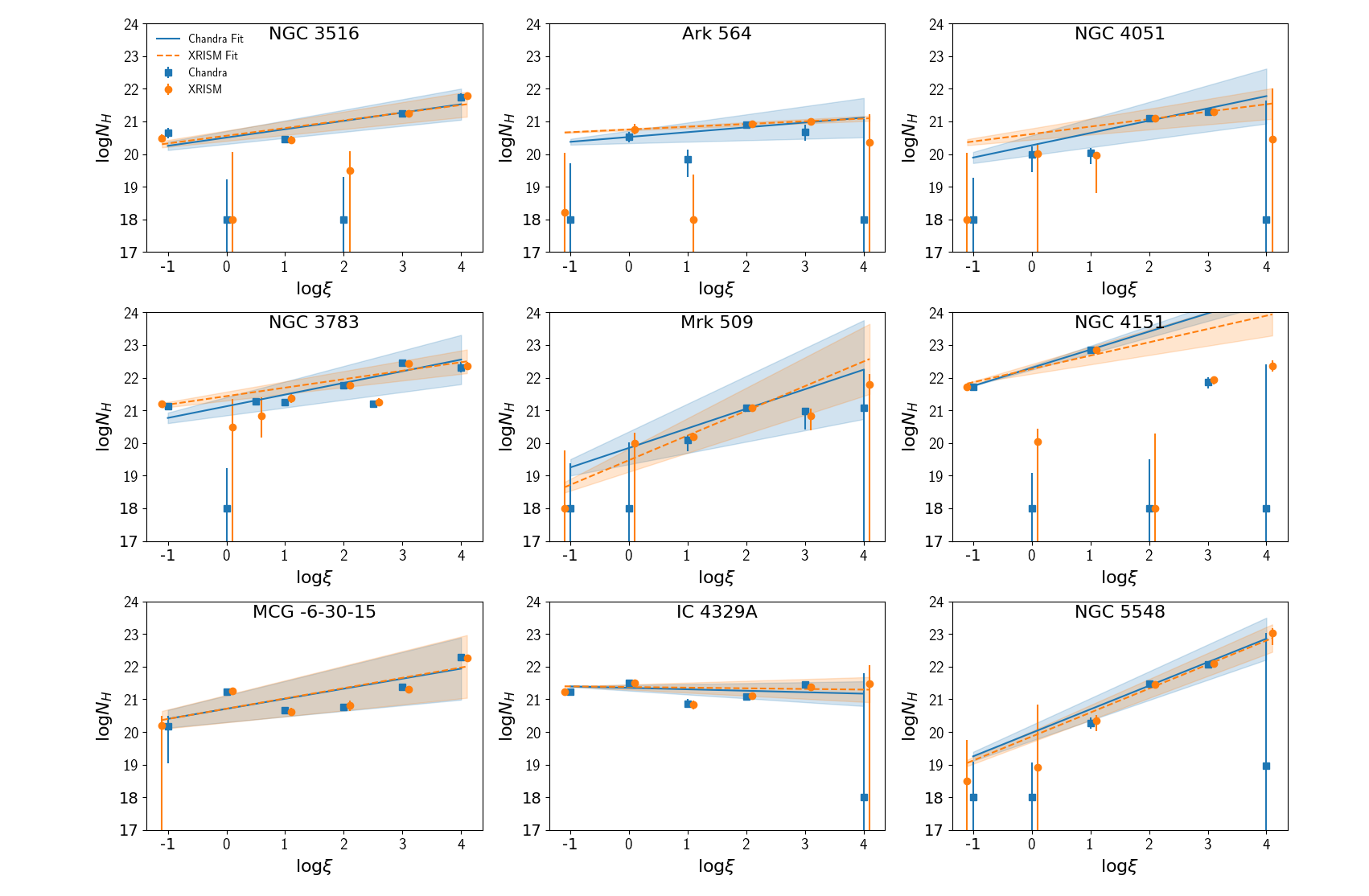}
    \caption{AGN AMDs of the present sample, for both HETG and \textit{Resolve}. Shaded regions represent the regression uncertainties. The fitted linear regression demonstrates The universal slowly increasing trend.}
    \label{fig:9AMD}
\end{figure}

Figure \ref{fig:9AMD} presents all the AMDs and their fitted slopes, for both spectrometers, and including $N_{\rm H}(\xi _i)$ uncertainties.
It can be seen from the figure and Table\,\ref{tab:columns} that the highest column densities occur in the high-$\xi_i$ components. In some AGNs it is difficult to identify these components using the HETG observations, therefore \textit{Resolve} spectra could have a meaningful advantage. 
The values of AMD slopes are detailed in Table\,\ref{tab:slopes}. 
The HETG slopes range between 0.00 - 0.72, and between --0.07 - 0.85 including the uncertainties.
Apparently, a slowly increasing AMD is a common property of Seyfert outflows. Apart from Mrk\,509, all slopes are tightly constrained to $\sim 0.1$ or better. 
The anomalously high \NHtot\ in NGC\,4151 is due to its intermittent obscuration (see Section \ref{sec:intro}). Its high $N_{\rm H}(\xi _i) > 10^{22}$\,cm$^{-2}$ values at $\log \xi = -1$ and $\log \xi = 1$, which none of the other AGNs features (see Table\,\ref{tab:columns}) likely have little to do with the outflow.

\begin{deluxetable}{lccc}
\tabletypesize{\scriptsize}
\tablewidth{0pt}
\tablecaption{AMD slope values}
\label{tab:slopes}
\tablehead{
\colhead{AGN} & \colhead{HETG} 
& \colhead{\textit{Resolve}}
}
\startdata
{NGC\,3516}     &  $0.26\pm0.07$  &  $0.24\pm0.05$ \\ 
{NGC\,3783}     &  $0.36\pm0.12$  &  $0.26\pm0.06$ \\
{MCG\,-6-30-15} &  $0.31\pm0.14$  &  $0.32\pm0.13$ \\
{Ark\,564}      &  $0.15\pm0.10$  &  $0.08\pm0.01$ \\
{Mrk\,509}      &  $0.60\pm0.25$  &  $0.76\pm0.18$ \\
{IC\,4329A}     &  $0.00\pm0.07$  &  $0.00\pm0.07$ \\
{NGC\,4051}     &  $0.38\pm0.13$  &  $0.23\pm0.07$ \\
{NGC\,4151}     &  $0.56\pm0.06$  &  $0.41\pm0.12$ \\
{NGC\,5548}     &  $0.72\pm0.10$  &  $0.73\pm0.06$ \\
\enddata
\end{deluxetable}

The consistency 
between AMD slope values of HETG and \textit{Resolve} (Table\,\ref{tab:slopes}) implies
one can accurately constrain \NH $(\xi_i)$ and reconstruct the AMD based on \resolve , with observation times half as long as those of \hetg , or shorter. 
Our results in Table\,\ref{tab:columns} show that \resolve\, has the sensitivity to measure \NH $(\xi_i=10^4)$ down to $\sim10^{21}$\,cm$^{-2}$, with a 100 ks exposure. Column densities of lower-$\xi_i$-components can even be constrained as low as $\sim10^{20}$\,cm$^{-2}$.

\begin{figure}
    \centering
    \includegraphics[width=\textwidth]{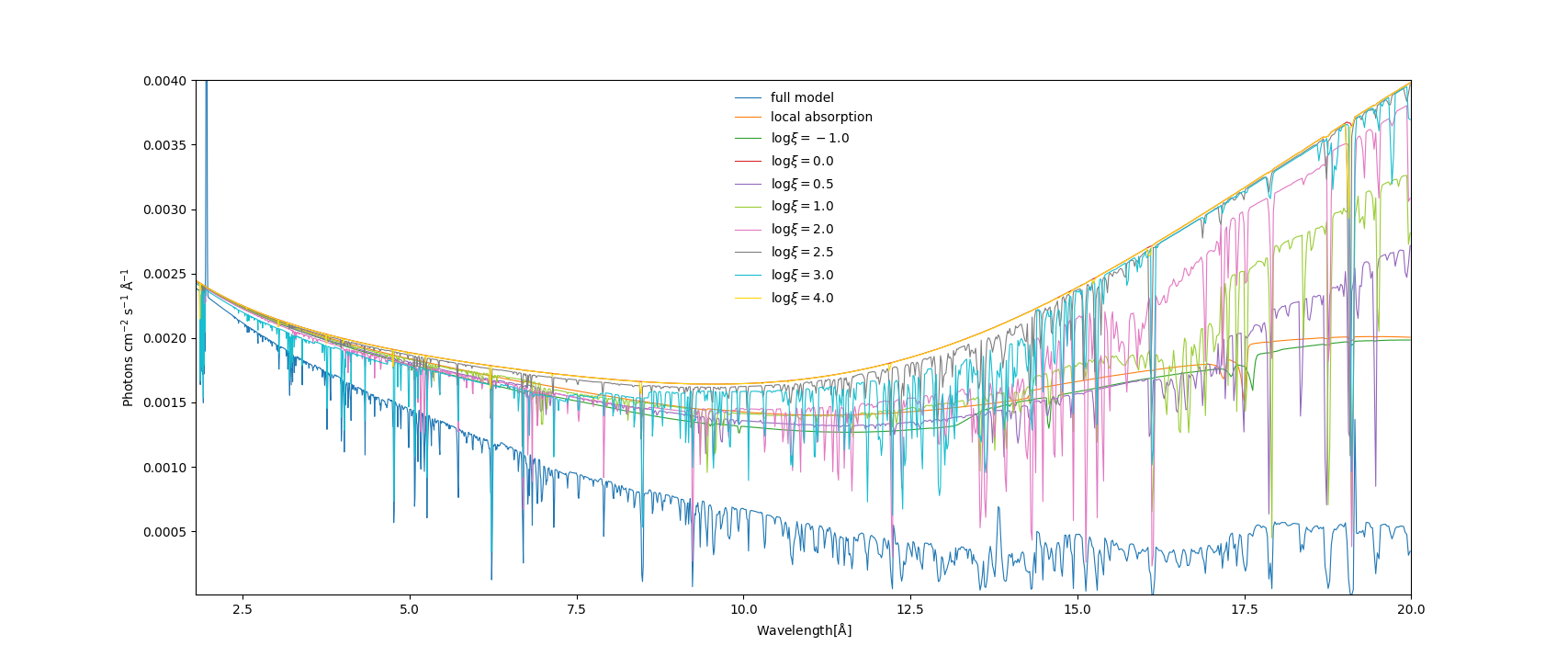}
    \caption{The absorption components of NGC\,3783 plotted separately.}
    \label{fig:3783model}
\end{figure}

In order to demonstrate the effect of the different $\xi_i$-components on the XSTAR based model, we plot each of them individually in Figure\,\ref{fig:3783model}, for the model of NGC\,3783 which has an abundance of resolved features. The Figure reveals several attributes of the model. The lower $\xi_i$-components absorb mainly the continuum, without many absorption lines. This is most evident for $\log\xi=-1$. Higher $\xi$ values absorb the continuum less and less; $\log\xi=4$ absorbs virtually no continuum. The column density of each $\xi_i$-component is predominantly determined by the imprint of its continuum slope (Figure\,\ref{fig:3783model}).
The Fe-M UTA, at $16-17$ \AA\ appears mainly in the $\log\xi=1$ component. Apparently, this XSTAR model does not fit the conspicuous UTA of NGC\,3783 properly, see residuals in Figure\,\ref{fig:3783spectra}. Since there is overlap in the lines between components, there is much freedom for the fit to lower or raise column densities of adjacent $\xi_i$-components, which is reflected in the uncertainties. The main driver of C-stat minimization is therefore the continuum. 

\subsection{AMD and other AGN Parameters}
 
\begin{figure}
    \centering
    \includegraphics[width=\textwidth]{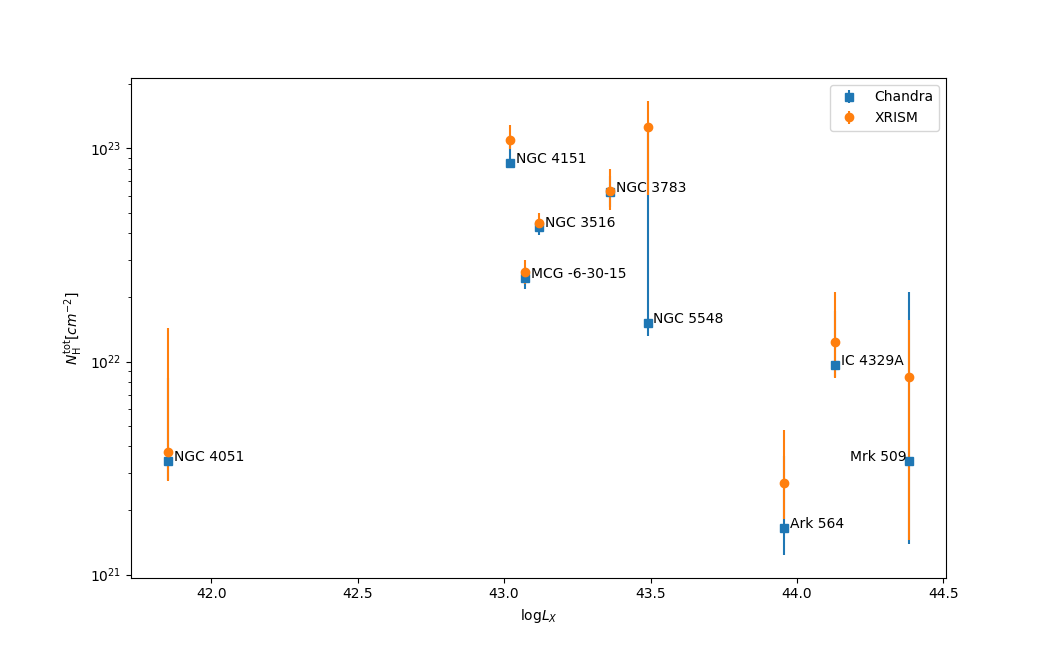}
    \caption{Total outflow column density vs. X-ray luminosity. The anti-correlation is apparent, with NGC\,4051 as an outlier.}
    \label{fig:L_totabs}
\end{figure}

\begin{figure}
    \centering
    \includegraphics[width=\textwidth]{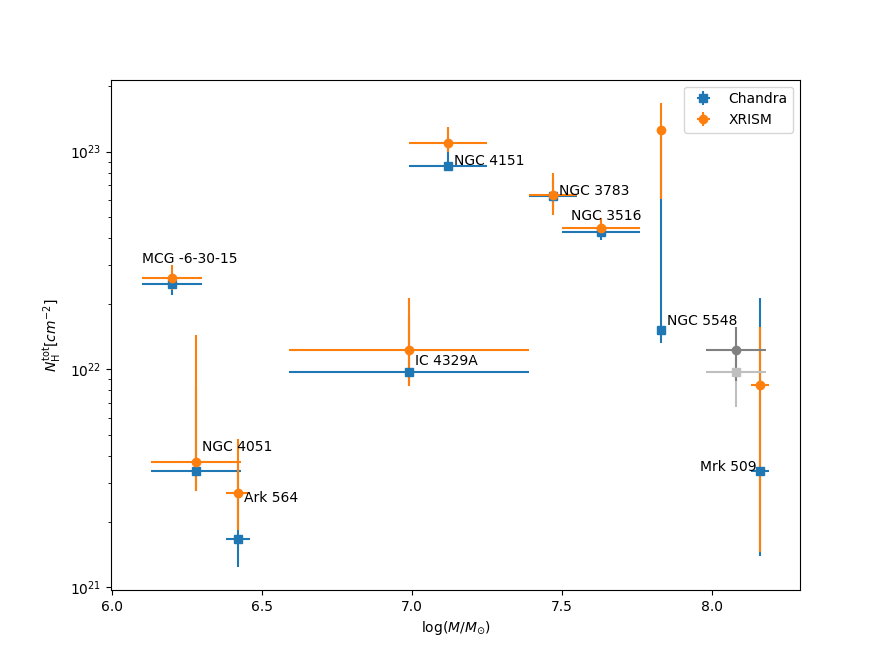}
    \caption{Total outflow column density vs. black hole mass. No obvious relation is seen. The grey dots represent the higher mass of IC\,4329A, see footnote f in Table \ref{tab:parameters}.}
    \label{fig:mass_totabs}
\end{figure}

\begin{figure}
    \centering
    \includegraphics[width=\textwidth]{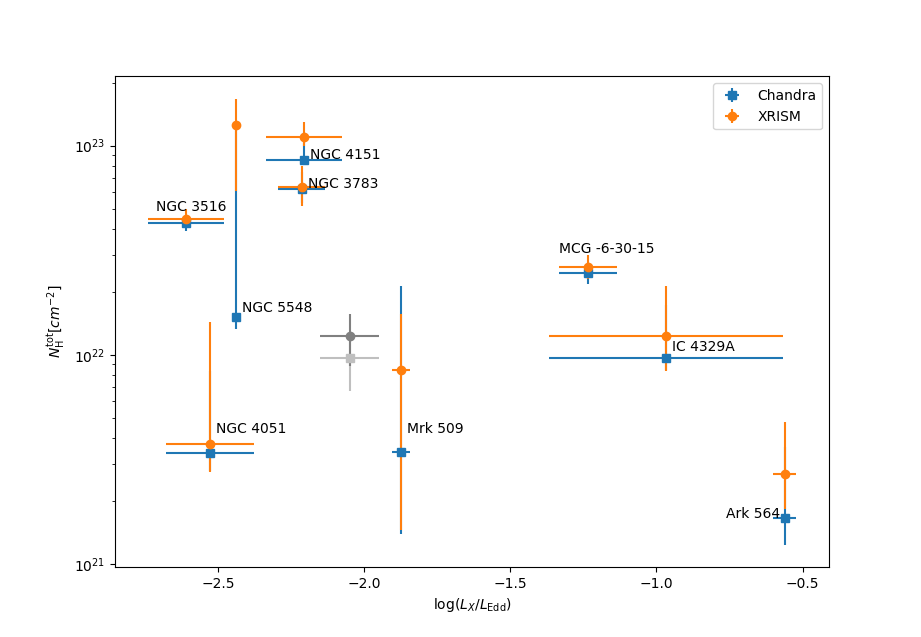}
    \caption{Total outflow column density vs. accretion efficiency. No obvious relation is seen. The grey dots represent the higher mass of IC\,4329A, see footnote (\textit{f}) in Table \ref{tab:parameters}.}
    \label{fig:ratioL_totabs}
\end{figure}

It remains to be understood what drives the outflow properties. We examine the connection between the AGN physical properties and \NHtot, meaning the sum of column densities of all $\log\xi_i$ components. In Figure \ref{fig:L_totabs} we show a relation between $\log$\NHtot\ and $\log L_{\rm X}$, which appear to be anti-correlated. The Pearson's correlation coefficient $r$ and Spearman's rank coefficient $r_s$ are -0.22 and -0.38, and p-values 0.56 and 0.31, respectively. However, by removing NGC\,4051, the anti-correlation improves dramatically (-0.83 and -0.83, with p-values 0.01 and 0.01, respectively). 
Note that the two main groups of AGNs in Fig.\,\ref{fig:L_totabs} differ by their $\log \xi = 4$ column density (high \NHtot\ low $L_{\rm X}$) or lack thereof (low \NHtot\ high $L_{\rm X}$). This separation may turn out to be a smooth transition, once \resolve\ better constrains this component.
Conversely, there seems to be no clear relation between \MBH\ and \NHtot , as can be seen in Figure \ref{fig:mass_totabs}. The coefficients there are $r = 0.17, r_s = 0.22$, with p-values 0.66 and 0.58, respectively.
Therefore, there is also no clear relation between \NHtot\ and $L_{\rm X}/L_{\text Edd} (\sim L_{\rm X}/$ \MBH ), see Figure \ref{fig:ratioL_totabs}. 
The coefficients there are $r = -0.40$, $r_s = 0.36$,  with p-values 0.29 and 0.36, respectively.
We also examine the connection between the AMD slope and the above AGN parameters. Since the slopes span a narrow range, we do not expect to find a strong relation. Indeed, in both cases we find no significant relation between the AMD slope and these AGN parameters. 

\section{Discussion} \label{sec:discussion}
Using archival \hetg\ grating observations of nine AGNs we constructed the AMDs of their outflows using at least 6 pre-defined $\xi_i$-components ranging over $-1 < \log \xi_i < 4$. Mrk\,509 requires three components, while the other three only provide upper limits, while NGC\,3783 requires seven components and one upper limit. This is a somewhat broader range than reported by \citet{McKernan2007}. The reason is that we assumed all pre-defined components, while \citet{McKernan2007} sought the minimal number of components that provided a satisfactory fit. 

The best-fit slopes of the various AMDs in the \hetg\ spectra span a range of 0.00 - 0.72 (-0.07 - 0.85 with uncertainties), which is consistent with the range of 0.0 -- 0.4 reported in \citet{Behar2009}.
\citet{Steenbrugge2005B} found a slope of $0.40\pm0.05$ for the outflow of NGC\,5548, which is inconsistent with our results. This may be due to the fact we used later observations of the outflow as well as the ones used in \citet{Steenbrugge2005B}.
In a comprehensive linear regression for the $\xi$-components of 17 different Seyfert outflows (including all of the present ones except NGC\,4151), \citet{Laha2014} find an AMD slope of $0.31\pm0.06$, which is consistent with the present slopes.
This suggests a ubiquitous AMD shape for AGN outflows of a shallow positive slope ($< 0.72$).

The present slope range of 0.00 - 0.72, corresponds in the MHD self similar solutions of $B\sim r^{q-2}$ \citep{Fukumura2010}, to $q = 0.79 - 1.0$, or approximately $B\sim 1/r$ in all outflows. Our results are marginally consistent with those of \citet{Stern2014}, $a\sim0.03$. \resolve\ spectra will allow us to measure the high-$\xi$ components with smaller uncertainties,  providing a more definitive AMD slope value.
The MHD outflow model is scalable with $q$, but other models will be confronted with these refined AMDs.


The anomalous high column densities of NGC\,4151  (\NH$(\xi_i)>10^{22}$\,cm$^{-2}$) in low-$\xi$  components are greater than those of any other AGN (Fig.\,\ref{fig:9AMD}).
The soft X-rays of NGC\,4151 are often heavily absorbed \citep{George1998,2005ApJ...633..693K} by K-edges of light elements that are likely not related to the steady outflow.
We analyzed the 2002 spectra where NGC\,4151 was in a high flux state.
Nevertheless, residual continuum absorption results in these high \NH$(\xi_i)$ values for low-$\xi_i$ components, although there are no clear absorption lines above 10\,\AA\ \citep[see also][]{2005ApJ...633..693K}. 

Previous works \citep{Holczer2007,Laha2014} find a gap in the AMD, between $\log \xi = 0.5 - 1.5$, and suggest this could be a universal feature due to thermal instability. 
\citet{Waters2021} show that in thermally driven winds, the buoyancy of gas clumps, and their disintegration within thermally unstable regions can remove this gap from the AMD.
The present method of a rigid $\xi$-grid does not provide unambiguous evidence for this gap, although marginal evidence can be seen in the AMDs of NGC\,3516, NGC\,3783, NGC\,4151 (Fig.\,\ref{fig:9AMD}).

Since the AMD slopes are relatively similar between AGNs, they point to a basic physical attribute of the outflows which is universal. On the other hand, the large dispersion in \NHtot, allows us to correlate it with the AGN fundamental properties. We find that the $\log$\NHtot plausibly anti-correlates with $\log L_{\rm X}$. A similar anti-correlation is found in the SUBWAY quasar sample between \NH$_{\text{I}}$ and $L_{\text{bol}}$ (Mehdipour et al., in preparation). Conversely, there is no correlation between \NHtot\ and \MBH\ or $L_{\rm X}/L_{\text{Edd}}$. Radiatively driven winds are actually expected to drive more mass with luminosity. 
The anti-correlation with $L_X$ thus might suggest that the X-ray flux moves gas out of the line-of-sight, leading to lower column densities. An alternative explanation is that high-$L_X$ AGNs totally ionize the wind, thus hiding its most ionized components. However, the similar AMDs of all AGNs and specifically the lack of increasing AMD slope with luminosity, suggests that outflows of low-$L$ sources are as ionized as those of high-$L$ ones.  
\citet{Blustin2005} found a possible, weak correlation between \NHtot\ and the bolometric luminosity \citep[cf. Fig. 5 in][]{Laha2016}, which hangs on four luminous quasars. Two of them, PG0844+349 and PG1211+143, have ultra-fast velocities, quite different from the Seyfert outflows \citep[e.g.,][]{Laha2014}. The two other sources, IRAS\,3349+2438 \citep[$L_{\rm X} \sim 6\times10^{44}$\,erg\,s$^{-1}$ and \NHtot  $= 1.2\pm0.3 \times10^{22}$ cm$^{-2}$,][]{Holczer2007}, and 
MR\,2251-178  \citep[$L_{\rm X} = 1.7-5.2 \times10^{44}$\,erg\,s$^{-1}$ and \NHtot  $= 3.2 - 6.3 \times10^{21}$ cm$^{-2}$,][]{Kaspi2004},  would strengthen the anti-correlation of Fig.\,\ref{fig:L_totabs}.


\section{Conclusions} \label{sec:conclusions}

Following the uniform analysis of a sample of nine Seyfert outflows we reach the following conclusions: 

\begin{itemize}
  \item The AMD slope, a proxy of the ionization distribution in the outflow is relatively flat. This slope is found to be a universal characteristic of the outflows, indicating a common wind-launching mechanism, or micro-physics that is not related to global properties of the AGN.
  \item The log of the total column density in the outflow \NHtot\ anti-correlates with the log of the X-ray luminosity $L_{\rm X}$, perhaps indicating that high-$L_{\rm X}$ sources clear absorbing gas from the line of sight.
  \item \resolve\ is expected to complement the AMDs of \hetg , with lower exposure times, by better constraining the high-velocity high-ionization components.
  This will provide a stricter confrontation of the measured AMDs (slopes) with the various theoretical models. 
  \item The present method of global fitting is insensitive to individual lines and suffers from possible limitations of the atomic data in the model. These will be further bench-marked with \resolve.
\end{itemize}

%

\vspace{5mm}
\facilities{
CXO, XRISM}


\software{Xspec \citep{Arnaud1996}, XSTAR \citep{Bautista2001}
          }

\begin{acknowledgements}
This work was supported by a Center of Excellence of The Israel Science Foundation (grant No. 2752/19) and performed in part at the Aspen Center for Physics, which is supported by National Science Foundation grant PHY-1607611.
\end{acknowledgements}

\bibliography{sample631}{}
\bibliographystyle{aasjournal}



\end{document}